\documentclass[prl,twocolumn,superscriptaddress,floats,showpacs]{revtex4}

\usepackage[dvips]{graphicx}

\usepackage{amsmath}

\usepackage{dcolumn}

\usepackage{rotating}

\newlength{\La} 
\newlength{\Lb} 
\newlength{\Lc} 
\newcommand{\grad}{\ensuremath{^\circ}}

\makeatletter
\DeclareRobustCommand{\chemical}[1]{%
  {\(\m@th
   \edef\resetfontdimens{\noexpand\)%
       \fontdimen16\textfont2=\the\fontdimen16\textfont2
       \fontdimen17\textfont2=\the\fontdimen17\textfont2\relax}%
   \fontdimen16\textfont2=2.7pt \fontdimen17\textfont2=2.7pt
   \mathrm{#1}%
   \resetfontdimens}}
\DeclareRobustCommand{\bchemical}[1]{%
  {\(\m@th
   \edef\resetfontdimens{\noexpand\)%
       \fontdimen16\textfont2=\the\fontdimen16\textfont2
       \fontdimen17\textfont2=\the\fontdimen17\textfont2\relax}%
   \fontdimen16\textfont2=2.7pt \fontdimen17\textfont2=2.7pt
   \mathbf{#1}%
   \resetfontdimens}}
\makeatother

\newcommand{\casrruo}{\chemical{Ca_{2-x}Sr_xRuO_4}}
\newcommand{\bcasrruo}{\bchemical{Ca_{2-x}Sr_xRuO_4}}
\newcommand{\srruo}{\chemical{Sr_2RuO_4}}
\newcommand{\caruo}{\chemical{Ca_2RuO_4}}
\newcommand{\bcaruo}{\bchemical{Ca_2RuO_4}}
\newcommand{\caruoen}{\chemical{Ca_{1.9}Sr_{0.1}RuO_4}}
\newcommand{\bcaruoen}{\bchemical{Ca_{1.9}Sr_{0.1}RuO_4}}
\newcommand{\caruoea}{\chemical{Ca_{1.8}Sr_{0.2}RuO_4}}

\newcommand{\caruoef}{\chemical{Ca_{1.5}Sr_{0.5}RuO_4}}

\begin{document}


\title{ Structural and magnetic aspects of the
metal insulator transition in \bcasrruo}

\author{O.~ Friedt}
\affiliation{Laboratoire L\'{e}on Brillouin,
C.E.A./C.N.R.S., F-91191-Gif-sur-Yvette CEDEX,
France}

\author{M.~ Braden}
\affiliation{Laboratoire L\'{e}on Brillouin,
C.E.A./C.N.R.S., F-91191-Gif-sur-Yvette CEDEX,
France}
\affiliation{Forschungszentrum Karlsruhe, IFP, Postfach 3640, D-76021
Karlsruhe, Germany}

\author{G.~ Andr\'{e}}
\affiliation{Laboratoire L\'{e}on Brillouin,
C.E.A./C.N.R.S., F-91191-Gif-sur-Yvette CEDEX,
France}

\author{P.~ Adelmann}
\affiliation{Forschungszentrum Karlsruhe, IFP, Postfach 3640, D-76021
Karlsruhe, Germany}

\author{S.~ Nakatsuji}
\affiliation{Department of Physics, Kyoto University,
Kyoto 606-8502, Japan}

\author{Y.~ Maeno}
\affiliation{Department of Physics, Kyoto University,
Kyoto 606-8502, Japan}
\affiliation{CREST, Japan Science and Technology Corporation, Kawaguchi,
Saitama 332-0012, Japan}


\begin{abstract}
The phase diagram of \casrruo\ has been studied by neutron
diffraction on powder and single-crystalline samples.
The experiments reveal antiferromagnetic order and structural
distortions characterized by tilts and rotations of the
RuO$_6$-octahedra.
There is strong evidence that the structural details of the isovalent
samples tune the magnetic as well as the electronic behavior. In
particular we observe for low Sr-concentration a
metal insulator transition associated with a structural change
and magnetic ordering.

\end{abstract}

\maketitle

\section{INTRODUCTION}

Layered perovskite ruthenates have attracted considerable interest
since the discovery of superconductivity in \srruo, which remains
till today the only known superconductor isostructural to the
cuprates \cite{maeno}. It is, therefore, expected that this
material can give further insight into the mechanism of High
Temperature Superconductivity. However, the origin of the spin-triplet pairing in \srruo\ \cite{ishida}is
far from understood. There is reasonable evidence that in this
material a coupling between electrons and magnetism is essential:
for example magnetic susceptibility \cite{neumeier} and low
temperature specific heat \cite{cp} exhibit similar enhancements.
It has been suggested that ferromagnetic fluctuations are
dominating the interaction leading to an unconventional pairing of
p-wave symmetry \cite{4}. This proposal was mainly inspired by the
fact that the perovskite SrRuO$_3$ is indeed an itinerant
ferromagnet \cite{srruo3}. The substitution of Sr by Ca in the
layered compound yields rather different physical properties.
First \caruo\ is an insulator at low temperatures
\cite{satoru-cro} and second it orders antiferromagnetically
\cite{7,8}, which clearly indicates that considering \srruo\ as
being close to ferromagnetic order is an oversimplification. More
recently band structure analyses on \srruo\ have predicted that
the magnetic susceptibility presents dominating peaks at
incommensurate wave vector positions, $q_0=(2\pi/3a, 2\pi/3a, 0)$,
which arise from Fermi-surface nesting \cite{mazin}. Inelastic
neutron scattering studies have perfectly confirmed these features
\cite{sidis}. In order to get an insight to the relation between
these incommensurate peaks in the \srruo\ susceptibility and the
antiferromagnetic order in \caruo\ it appears essential to study
the entire phase diagram of \casrruo.

The physics of \casrruo\ attracts interest not only in the context
of the superconductivity in \srruo. In our first paper we have
demonstrated that in a \caruo\ sample containing excess oxygen
(Ca$_2$RuO$_{4.07}$, O-Ca$_2$RuO$_4$),
one finds a structural transition from a metallic
high-temperature phase into a non-metallic distorted
low-temperature phase hence a metal-insulator transition \cite{7}.
The observation of the antiferromagnetic order in the non-metallic
phase suggests the interpretation that this transition is of the
Mott-type. The high temperature metallic phase of O-\caruo\ is
characterized by an octahedron shape almost identical to that
observed in \srruo, whereas the in-plane Ru-O bond lengths are
significantly enhanced in the low temperature insulating phase.
In addition there is a stronger tilt of the RuO$_6$ octahedra in the
low temperature phase. The structural transition is of
the first order type, as seen in the large hysteresis with
coexistence of the two phases, and presents a lattice expansion of
$\frac{\Delta V}{V}\sim1\,\%$ upon cooling. 
Recently Alexander et al. have found a sudden increase in the
resistivity of stoichiometric Ca$_2$RuO$_4$ at 357\ K 
\cite{alexander}, suggesting
that the metal insulator transition seen in O-Ca$_2$RuO$_4$
occurs just at higher temperature in the stoichiometric compound.
Nakatsuji et al. have revealed the entire magnetic phase diagram of \casrruo, and found the 
metal insulator transition by resistivity
and magnetic susceptibility measurements \cite{nakatsuji-sces,nakatsuji-mag}.
They observe that the anomalies in the resistivity 
are rapidly shifted to lower temperature for increasing Sr-content,
for x$>$0.2 samples stay metallic till low temperature.
Also Cao et al. report a decrease of the metal insulator transition temperature upon Sr-doping; 
in addition they observe the similar suppression for La-doping \cite{cao-2000}.
However, no diffraction study on structural aspects of the metal insulator transition as function of 
doping has been reported so far.

In many aspects the
metal insulator transition in  \casrruo\ resembles that in
V$_2$O$_3$ \cite{mott,bao}; the simpler crystal structure in the case
of the ruthenate should be favorable for the analyses and their
understanding.

We have extended our previous diffraction studies to the
entire Sr-content range in \casrruo\ proving that the structural
distortion accompanying the metal insulator transition persists to
x$\sim$0.15 in O-stoichiometric samples. For higher Sr-content we
find a distinct crystal structure, space group I4$_1$/acd, which
again presents a structural phase transition upon cooling,
however, between two metallic phases.

\section{EXPERIMENTAL}
The stoichiometric sample of \caruo\ (S-\caruo) used already in
our previous work has been further analyzed at higher
temperatures. In addition we have prepared samples of \casrruo\
with x=0.1, 0.2, 0.5, 1.0  and 1.5 by the technique described in Ref.
\cite{satoru-cro}, details will be given else-where \cite{satoru-neu}. 
Thermogravimetric studies indicate a
stoichiometric oxygen content in these mixed compounds. All
samples were characterized by x-ray diffraction and by magnetic
measurements.

\begin{figure}[!ht]
\begin{center}
\includegraphics*[width=0.9\columnwidth]{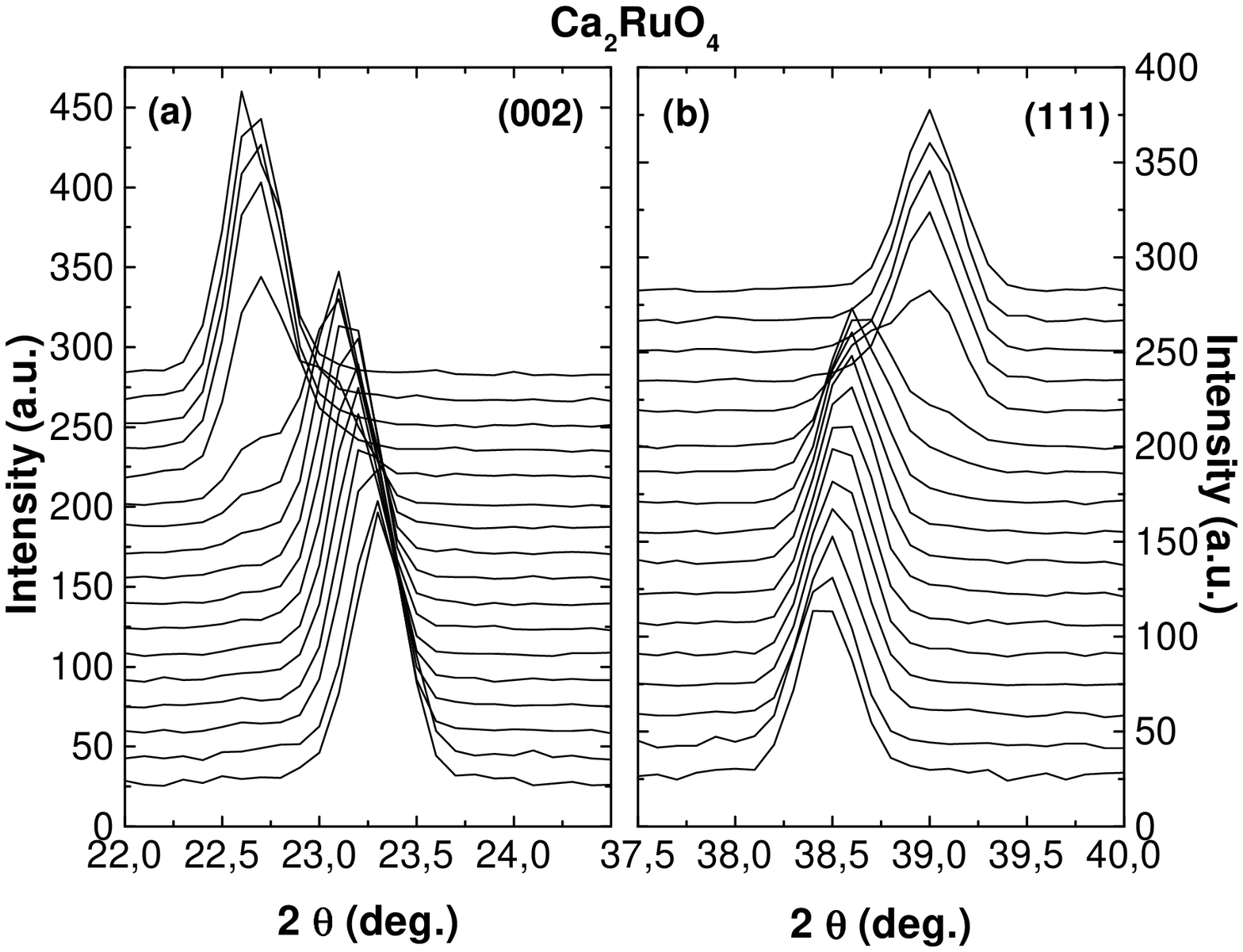}\\
\includegraphics*[width=0.9\columnwidth]{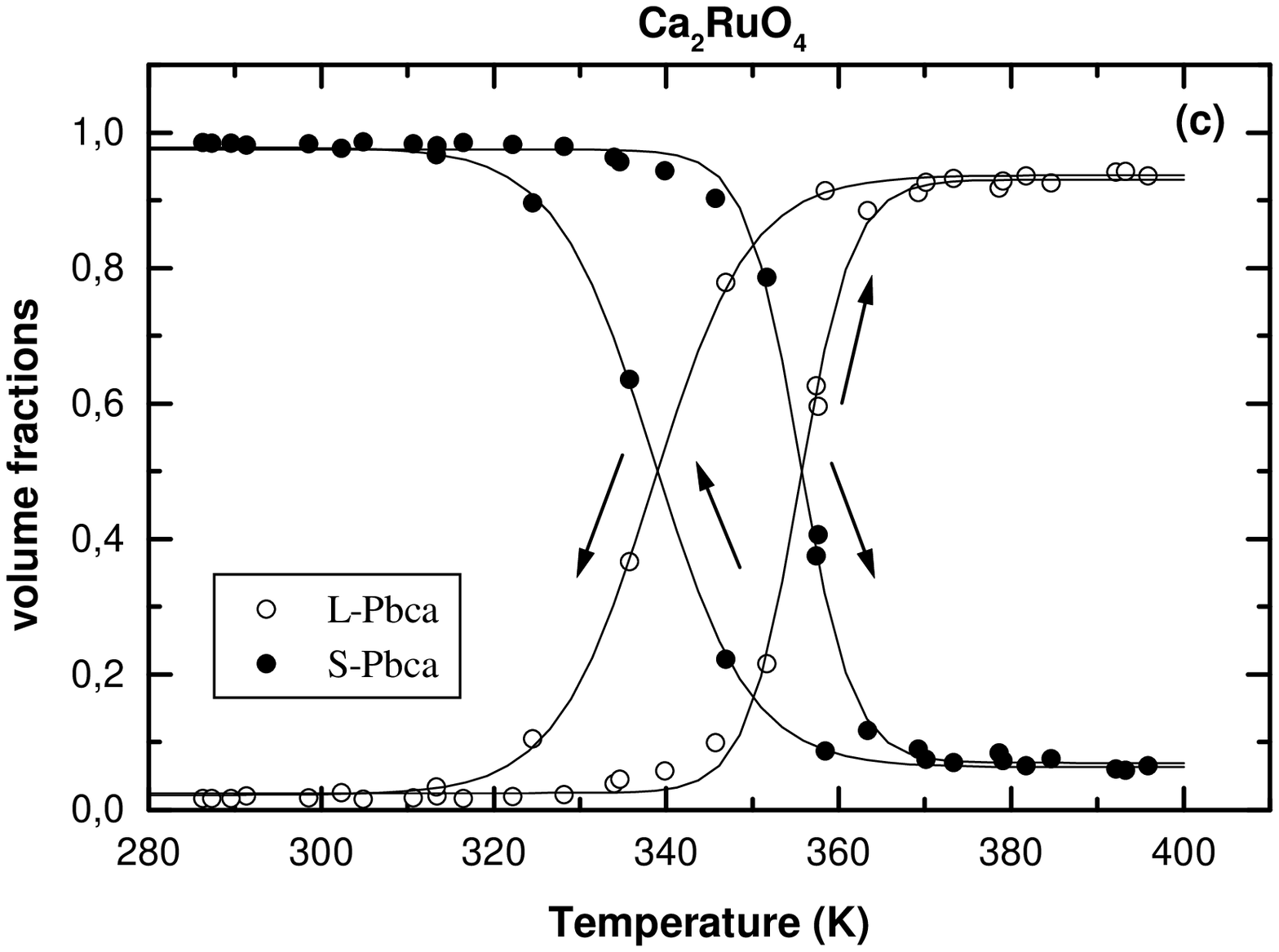}%
\caption{Comparison of the high flux diffraction patterns, showing the (002) (a) and (111) (b) reflections, obtained in stoichiometric \caruo\ at temperatures between 286 and 400\, K.
The lowest lines correspond to T=286\, K and the following one approximately
to 5\, K steps. (c) Temperature dependence of the partial volume fractions of the L-Pbca and S-Pbca phases for \caruo. The temperature variation is indicated by arrows.}
\label{fig:1}
\end{center}
\end{figure}

Neutron diffraction studies were performed at the ORPHEE reactor using the
two diffractometers 3T.2 ($\lambda =1.226\,$\AA) and 
G4.1 ($\lambda =2.43\,$\AA).
With the shorter wavelength instrument it is possible to perform complete
Rietveld structure analyses whereas the longer wavelength multi-counter
machine offers a high flux permitting the measurement of temperature
dependencies.  For more details of the diffraction analyses see our first
paper \cite{7}.

Two single-crystals with x=0.2 and x=0.5 were obtained by a
floating zone technique; these were examined on the
two-axis diffractometer 3T.1 using pyrolitic graphite
monochromator and filters.

\section{Results and discussion}

\subsection{Metal insulator transition in S-\bcaruo\ and in \bcaruoen}
The strong temperature dependence of the structural parameters of
S-\caruo\ observed near room temperature indicates a structural
phase transition in the temperature range 350--400\, K. Indeed such
a transition has been recently reported by Alexander et al. based
on x-ray diffraction and resistivity studies \cite{alexander}. In
a first view, one might argue that the transition in O-\caruo\ is
just shifted to higher temperature in the stoichiometric sample,
but the detailed structure analysis presents significant
differences.

Using a cryo-furnace the structure was studied on G4.1 by 
recording a complete hysteresis cycle. At temperatures near 340\, K, we already observe
two phases, the low-temperature phase is characterized by a small
c-lattice parameter (S-Pbca) compared to the high temperature phase
with long c (L-Pbca);
the averaged in-plane parameter exhibits the opposite behavior as can be seen
in Fig.~\ref{fig:1} (b). The transition in stoichiometric \caruo ~
is sharper than in O-\caruo; at 365\, K
it is almost complete, and at 395\, K there is no sign of the low
temperature S-Pbca phase in the high flux patterns, parts of the
high flux patterns are shown in Fig.~\ref{fig:1} (a),(b). To avoid possible variation in oxygen stoichiometry during the hysteresis cycle, the temperature was limited to 400\,K. Also the high resolution pattern obtained at 400\ K can be refined by a single
S-Pbca-phase, the results are given in Table \ref{tab:1}.
(Throughout the paper we use the same notations as in ref. \cite{7};
O(1) denotes the oxygen in the RuO$_2$-planes and O(2) the apical one.
The rotation angle is unique and designed by $\Phi$, whereas the tilt
angle may be determined at the two oxygen sites, 
$\Theta\text{-}O(1)$ and $\Theta\text{-}O(2)$;
in the notation used here the tilt is always around an axis 
close to the b-axis in Pbca.) In Fig.~\ref{fig:1} (c) we show the temperature dependence of 
the L- and S-Pbca-phase volumes. The transition temperature obtained in up-strike, $T_S$=356\, K, is in excellent agreement to that observed by Alexander et al. \cite{alexander}, $T_S$=357\, K.
The  observed hysteresis of about 20\, K clearly confirms the first order nature of 
the transition. We conclude that the combined electronic and structural transition first observed in O-\caruo\ also occurs in the stoichiometric compound.

In order to complete the study of the phase transition in
stoichiometric \caruo\  high resolution patterns have been
recorded at 400 and at 180\, K. The structure of the high-temperature
L-Pbca-phase could  not be established unambiguously in O-\caruo\ \cite{7}.
Space group Pbca combines a rotation of the octahedra around the c-axis
with a tilt around an axis parallel to an edge of the octahedron basal plane.
In contrast in space group P2$_1$/c the tilt can be about an arbitrary 
direction, in particular around a RuO-bond.
The difference in the two symmetries may be easily tested in the
O(1)-positions, the tilt around the Ru-O-bond displaces only
one O(1)-site parallel to c, whereas the tilt around an axis parallel
to the edge displaces all O(1)-positions about the same distance.
In other words, in Pbca all O(1)-sites are equivalent whereas there 
are two of them in P2$_1$/c.
In the cuprates the two tilt schemes correspond to the LTO and LTT phases,
see for example \cite{lasrcuo-struc}.
In the excess oxygen compound a better description of the data was obtained
in space group P2$_1$/c and in particular free refinement of the two distinct
O(1)-displacements along c led to significant difference 
with one  displacement almost vanishing, leading to a LTT equivalent tilt-scheme.
The structure was finally refined with one O(1)-site being fixed at $z$=0.
In case of the stoichiometric compound at 400\, K this is definitely not
an appropriate model.
In space  group Pbca one obtains a R-value of 5.39\% which increases to
5.62\% for the P2$_1$/c-model with one $z$-O(1) fixed to zero. 
Refining the P2$_1$/c
phase independently still gives a lower R-value than Pbca, but this
difference is not significant any more. Therefore we conclude
that the high temperature structure in stoichiometric \caruo\ has the
same space group Pbca as the low temperature structure. We differentiate 
these phases as L-Pbca and S-Pbca respectively due to their different
c-parameters. The large amount of excess oxygen ($\delta$=0.07
in Ca$_2$RuO$_{4+\delta}$) seems to be responsible
for the distinct diffraction pattern in the excess O compound.
\begin{sidewaystable*}[p]
\begin{center}
\begin{tabular}{l d d d d d d d d}\hline\hline
Composition & \multicolumn{1}{c}{x=0.0} & \multicolumn{1}{c}{x=0.0} & \multicolumn{1}{c}{x=0.1} & \multicolumn{1}{c}{x=0.1} & \multicolumn{1}{c}{x=0.2} & \multicolumn{1}{c}{x=0.2}  & \multicolumn{1}{c}{x=0.5} & \multicolumn{1}{c}{x=0.5}\\
Temperature & \multicolumn{1}{c}{180\, K} & \multicolumn{1}{c}{400\, K} & \multicolumn{1}{c}{10\, K} & \multicolumn{1}{c}{300\, K} & \multicolumn{1}{c}{10\, K} & \multicolumn{1}{c}{300\, K} & \multicolumn{1}{c}{10\, K} & \multicolumn{1}{c}{300\, K}\\
Space group & \multicolumn{1}{c}{$S\text{-}Pbca$} & \multicolumn{1}{c}{$L\text{-}Pbca$} & \multicolumn{1}{c}{$S\text{-}Pbca$} & \multicolumn{1}{c}{$L\text{-}Pbca$} & \multicolumn{1}{c}{$L\text{-}Pbca$} & \multicolumn{1}{c}{$L\text{-}Pbca$}  & \multicolumn{1}{c}{$I4_1/acd$} & \multicolumn{1}{c}{$I4_1/acd$}\\ \hline
$a$ (\AA) & 5.3945(2) &  5.3606(3) & 5.4207(4) &  5.3494(3) & 5.3304(4) & 5.3295(5) &  5.3195(1) &  5.3395(1)\\
$b$ (\AA) & 5.5999(3) &  5.3507(3) & 5.4802(6) &  5.3420(3) & 5.3190(5) & 5.3232(4) &  \multicolumn{1}{c}{\,\,\,\,\,\,\,``} & \multicolumn{1}{c}{\,\,\,\,\,\,\,``}\\
$c$ (\AA) & 11.7653(5) & 12.2637(4) & 11.9395(6) & 12.3219(4) & 12.4094(7) & 12.4506(7) & 2\cdot 12.5867(3) & 2\cdot 12.5749(3)\\
Vol (\AA$^3$) & 355.41(3) & 351.76(2) & 354.67(5) & 352.12(3) & 351.84(5) & 353.22(5) & 2\cdot 356.17(1) & 2\cdot 358.51(1) \\
$|\epsilon|$ & 0.0187 & 0.0009 & 0.0055 & 0.0007 & 0.0011 & 0.0006 & / & /\\
$R_{wp}$ (\% ) & 5.81 & 5.39 & 5.16 & 5.84 & 5.95 & 5.48 & 5.15 & 4.26 \\
 & & & & & & & & \\
Ca $x$& 0.0042(4) & 0.0110(5) & 0.0083(4) & 0.0099(6)  & 0.0133(7) & 0.0103(15) & 0 & 0 \\
\hspace*{\La}$y$ & 0.0559(4) & 0.0269(5) & 0.0425(4) & 0.0214(6) & 0.0169(9) & 0.0157(12) & \multicolumn{1}{c}{$\,\,\,1/4$} & \multicolumn{1}{c}{$\,\,\,1/4$} \\
\hspace*{\La}$z$ & 0.3524(2) & 0.3479(1) & 0.3505(2) & 0.3481(1) & 0.3475(2) & 0.3482(2) & 0.5492(1) & 0.5494(1)\\
U$_{iso}$ (\AA$^2$) & 0.0052(5) & 0.0112(4) & 0.0039(5) & 0.0053(6) & 0.0073(4) & 0.0138(9) & 0.0022(3) & 0.0090(3)\\
Ru U$_{iso}$ (\AA$^2$) & 0.0025(4) & 0.0050(4) & 0.0011(3) & 0.0045(4) & 0.0007(4) & 0.0029(5) & 0.0014(3)& 0.0042(3)\\
O(1) $x$  & 0.1961(3) & 0.1939(4) & 0.1974(3) & 0.1934(4) & 0.1958(6) & 0.1944(6)  & 0.1933(2) & 0.1949(2)\\
\hspace*{\Lb}$y$ & 0.3018(4) & 0.3064(4) & 0.3016(3) & 0.3056(4) & 0.3064(6) & 0.3070(6) & \multicolumn{1}{c}{$\qquad\,\mathnormal{x}+1/4$} & \multicolumn{1}{c}{$\qquad\, \mathnormal{x}+1/4$}\\
\hspace*{\Lb}$z$ & 0.0264(2) & 0.0147(2) & 0.0229(1) & 0.0127(2) & 0.0126(3) & 0.0075(6) & \multicolumn{1}{c}{$\,\,\,1/8$} & \multicolumn{1}{c}{$\,\,\,1/8$}\\
U$_{\bot-\text{plane}}$ (\AA$^2$) & 0.0047(5) & 0.0116(6) & 0.0024(6) & 0.0100(5) & 0.0023(9) & 0.0088(8) & 0.0052(16) & 0.0092(16) \\
U$_{\|-\text{plane}}$ (\AA$^2$) & 0.0055(5) &  0.0033(5) & 0.0031(6) &  0.0020(6) & 0.0005(2) & 0.0054(5) & 0.0024(8) & 0.0040(7)\\
U$_{\text{long-axis}}$ (\AA$^2$) & 0.0112(10) & 0.0205(8) & 0.0171(9) & 0.0245(12) & 0.0166(16) & 0.0228(18) & 0.0152(6) & 0.0211(6)\\
O(2) $x$ & -0.0673(3) & -0.0386(3) & -0.0583(2) & -0.0340(4) &  -0.0368(5) &  -0.0187(14) & 0 & 0\\
\hspace*{\Lc}$y$ & -0.0218(4) & -0.0067(4) & -0.0141(4) & -0.0079(6) & -0.0078(5) &-0.0051(10)& \multicolumn{1}{c}{$\,\,\,1/4$} & \multicolumn{1}{c}{$\,\,\,1/4$}\\
\hspace*{\Lc}$z$ & 0.1645(2) & 0.1656(1) & 0.1645(1) & 0.1649(1) & 0.1652(2) & 0.1644(2)& 0.4568(1) & 0.4568(1)\\
U$_{\bot}$ (\AA$^2$) & 0.0093(5) & 0.0169(6) & 0.0070(5) & 0.0151(5) & 0.0044(7) & 0.0162(10) & 0.0084(3) & 0.0142(3)\\
U$_{\|}$ (\AA$^2$) & 0.0042(8) & 0.0117(8) & 0.0053(9) & 0.0079(7) & 0.0035 & 0.0051(10) & 0.0050(7) & 0.0080(7)\\\hline
Ru-O(1) (\AA) & 2.004(2) & 1.949(2) & 1.987(2) & 1.939(2) & 1.927(3) & 1.928(4) & 1.929(1) & 1.933(1)\\
 & 2.018(2) & 1.950(2) & 1.988(2) & 1.948(2) & 1.942(3) & 1.937(3) &  / & /\\
Ru-O(2) (\AA) & 1.973(1) & 2.042(1) & 1.991(2) & 2.040(2) & 2.060(3) & 2.050(3) & 2.059(3) & 2.057(3)\\
Ru-O$_{\text{aver}}$ (\AA) & 1.998 & 1.980 & 1.989 & 1.973 & 1.976 & 1.972 & 1.972 & 1.974\\
O(1)-O(1)$\|$a (\AA) & 2.828(1) & 2.771(1) & 2.822(1) & 2.758(1) & 2.750(2) & 2.739(2) & 2.727(2) & 2.734(2)\\
O(1)-O(1)$\|$b (\AA) & 2.860(1) & 2.742(1) & 2.799(1) & 2.739(1) & 2.722(1) & 2.727(2) & \multicolumn{1}{c}{\,\,\,\,\,\,\,``} & \multicolumn{1}{c}{\,\,\,\,\,\,\,``}\\
Vol RuO$_6$ (\AA$^3$) & 10.64 & 10.34 & 10.48 & 10.27 & 10.28 & 10.21 & 10.21 & 10.25\\
$\Theta$-O(1) (deg) & 12.69(10) & 7.49(10) & 11.16(5) & 6.52(10) & 6.6(2) & 3.9(3) & / & /\\
$\Theta$-O(2) (deg) & 11.19(8) & 5.91(7) & 9.40(7) &  5.25(11) & 5.59(10) &  2.9(3) & / & /\\
$\Phi$ (deg) & 11.93(5) & 12.77(6) & 11.77(5) & 12.65(6) & 12.47(9) & 12.69(10) & 12.78(3) & 12.43(3)\\\hline\hline
\end{tabular}
\caption{Results of the high resolution powder diffraction
analyses on \casrruo\ at 300 and at 10\, K.
O(1) denotes the oxygen in the RuO$_2$-planes and O(2) the apical one.
The rotation angle is unique and designed by $\Phi$ whereas the tilt
angle may be determined at the two oxygen sites, $\Theta$-O(1) and $\Theta$-O(2).
Note that in Pbca the tilt is always around the b-axis, whereas the
orthorhombic splitting may change sign. U$_{\|}$ and U$_{\bot}$ denote the mean-square atomic displacements parallel and perpendicular to the Ru-O bonds.
O(1)-O(1)$\|$a,b denote the length of the edge of the 
octahedron basal plane along a,b respectively.
Errors are given in parentheses and result from the least square fitting;
therefore they do not take account of systematic errors and underestimate
effects of correlations.
} \label{tab:1}
\end{center}
\end{sidewaystable*}

\begin{figure}[!ht]
\begin{center}
\includegraphics*[width=0.9\columnwidth]{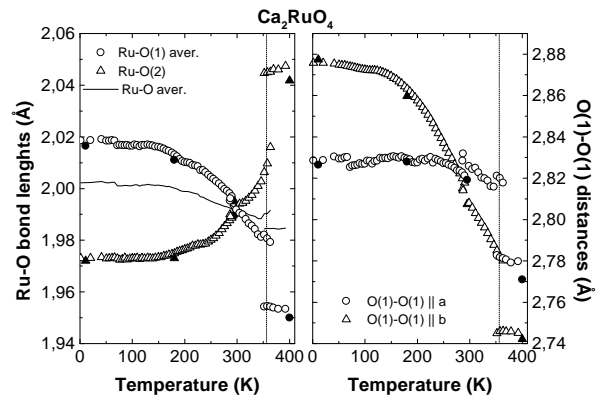}%
\caption{Temperature dependence of the RuO(1)-bond distances in
stoichiometric \caruo\ and that of the edge lengths of the
octahedron basal plane.} \label{fig:2}
\end{center}
\end{figure}

Combining the new and the previous results \cite{7} we get the
entire picture of the phase transition in S-\caruo.
Fig.~\ref{fig:2} presents the Ru-O bond-lengths, which show a discontinuous change at the
L-Pbca-S-Pbca transition: the
in-plane bonds become elongated and the out-of-plane one is
shrinking upon cooling. Close to the transition  -- but in the S-Pbca
phase --,  the octahedron is still elongated
along c in the temperature range 300--330\, K.
Upon further cooling one observes a continuous and even
larger change in the same sense: the octahedron at low temperature
is finally flattened along c. The edge lengths of the octahedron
basal plane also show a discontinuous effect at the transition
followed by a pronounced change in the S-Pbca phase. 
In accordance
to the elongation of the Ru-O(1)-bonds these lengths increase
upon cooling into the S-Pbca-phase where they are still split, the
octahedron basal plane is shorter along the tilt axis immediately 
below the transition. 
Upon further cooling, the ratio, however, becomes inversed, and the
basal plane is strongly stretched along the tilt axis. 

\begin{figure}[!ht]
\begin{center}
\includegraphics[width=0.9\columnwidth]{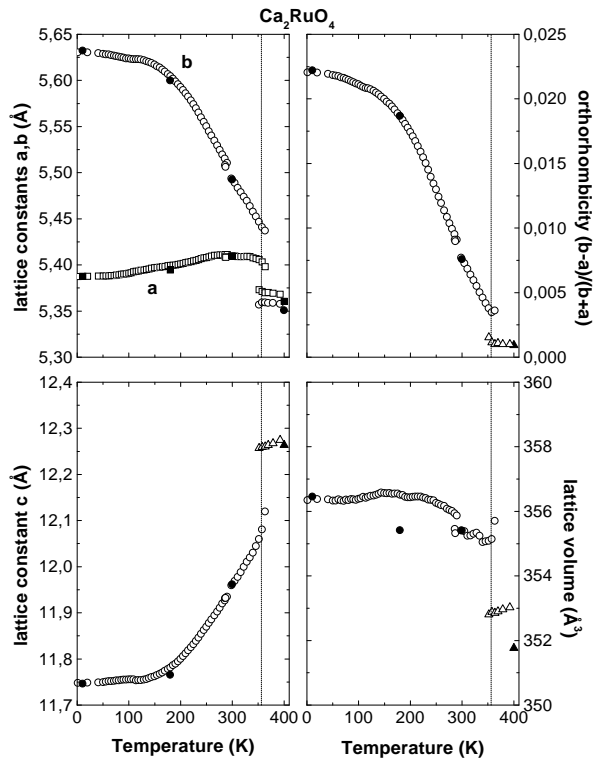}%
\caption{Temperature dependence of the lattice parameters in
stoichiometric \caruo. Open symbol designate the results
obtained from the high-flux patterns and closed symbols those
from the high resolution studies.} \label{fig:3}
\end{center}
\end{figure}

In  La$_2$CuO$_4$ \cite{lasrcuo-struc} and also in all L-Pbca phase ruthenates
studied here, the orthorhombic splitting is opposite to the expectation
of a rigid tilt, i.e. the lattice is longer perpendicular to the tilt-axis.
This effect originates from the forces in the La-O- or Sr-O-layer
where one distance strongly decreases due to the opposite displacements of 
an apical oxygen  and a neighboring La/Sr-site.
The stretching of the lattice perpendicular to the tilt axis reduces
the pronounced shrinking of this bond and may be seen in the orthorhombic
splitting. One should hence
consider this behavior as the normal one arising from
the structural arrangement.
Nevertheless the elongation of octahedron basal plane 
might influence the electronic and magnetic properties,
see the discussion below.
The orthorhombic lattice in \caruo\ is elongated along the tilt-axis
at low temperature much more than
what might be expected for a rigid 
octahedron tilt; the lattice constants are given in
Fig.~\ref{fig:3}. 
It is the stretching of the octahedra which causes this
orthorhombic splitting, and 
this behavior should be considered as highly anomalous.

The transition from L-Pbca to S-Pbca is further characterized by an
increase in the tilt angles, $\Theta$-O(1)($\Theta$-O(2)) increase from
7.5\grad\ (5.9\grad) at 400\, K to 11.2\grad\ (9.2\grad) at 295\, K.
Upon cooling in the S-Pbca phase these angles first continue to increase
till about 180\, K and are almost constant below, see Fig. 6 in ref. \cite{7}.
The rotation angle decreases by about one degree during the
transition into the S-Pbca phase and is constant over the whole 
S-Pbca temperature range.

\begin{figure}[!ht]
\begin{center}
\includegraphics[width=0.9\columnwidth]{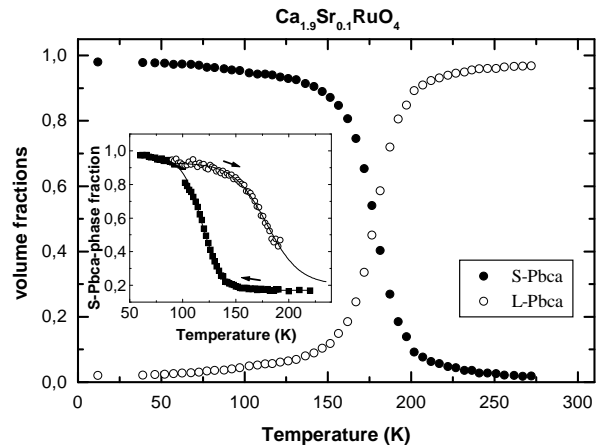}%
\caption{Temperature dependence of the volume fractions of the
L-Pbca and S-Pbca phases in \caruoen\ measured on the high flux
neutron diffractometer. The inset shows the hysteresis of the
transition as obtained by x-ray diffraction.} \label{fig:4}
\end{center}
\end{figure}

In \caruoen\ we find a structural transition similar to
that observed in \caruo. Fig.~\ref{fig:4} presents the volume
fractions of the L-Pbca and S-Pbca phases as a function of
temperature. The inset presents the hysteresis of the transition
which amounts to about 50\, K. For the discussion of any temperature
dependent property one has to take account of this hysteresis.

The high resolution room temperature data in the metallic L-Pbca phase
is again difficult to analyze.
Refinements in space group Pbca and in space group P2$_1$/c with
or without fixed O(1)-site are very close concerning their R-values.
However, like in S-\caruo ~-- and in contrast to O-\caruo ~-- the
P2$_1$/c fit does not suggest different O(1)-$z$-displacements.
Powder diffraction alone, however, is unable to differentiate
between Pbca and P2$_1$/c space groups. A single crystal
diffraction experiment on a single crystal of almost identical
composition, Ca$_{1.85}$Sr$_{0.15}$RuO$_4$, definitely
excludes the tilt around the RuO-bond \cite{braden-up}.
Since P2$_1$/c is a sub-group of Pbca, the description of
Bragg reflection data must be better due to the larger number
of free parameters. However an extremely small improvement of the
R-value, 5.065\%\ for the P2$_1$/c-model compared to 5.078 \%\  
for the Pbca-model,  does not support the symmetry reduction \cite{braden-up}.
Therefore, we feel confident to analyze the powder diffraction patterns
on excess oxygen free samples in the L-Pbca-structure.

\begin{figure}[!ht]
\begin{center}
\includegraphics[width=0.9\columnwidth]{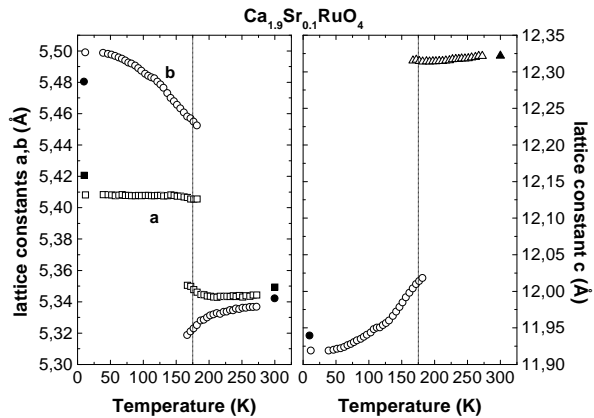}%
\caption{Temperature dependence of the lattice parameters in
\caruoen; open symbols designate the results obtained from the
high flux studies and closed symbols those from the
high resolution experiments.} \label{fig:5}
\end{center}
\end{figure}

The transition in \caruoen\ is much better defined than that in
O-\caruo\ and there is no L-Pbca phase remaining at low
temperature. Therefore, it has been possible to analyze the
structural details in this compound with the G4.1 data. Again high
resolution patterns have been recorded at 295 and at 11\, K with
the results of the refinements given in Table \ref{tab:1}; high
flux patterns were measured in temperature steps of 5\, K upon
heating. Fig.~\ref{fig:5} presents the temperature dependence of
the lattice parameters showing similar discontinuities at the
transition, T$_S\simeq$ 175\, K upon heating, as the stoichiometric sample.

\begin{figure}[!ht]
\begin{center}
\includegraphics[width=0.9\columnwidth]{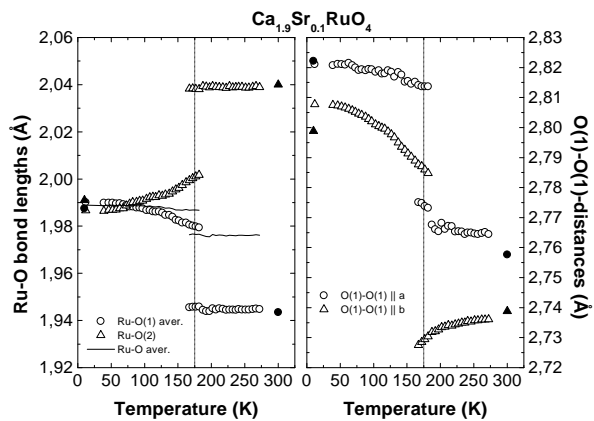}%
\caption{Temperature dependence of the Ru-O bond distances and
that of the octahedron basal-plane-edge lengths in \caruoen.}
\label{fig:6}
\end{center}
\end{figure}

In contrast to the stoichiometric compound, the temperature
dependence of the structural parameters within the 
S-Pbca phase is only small. 
In analogy to Fig.~\ref{fig:2}, Fig.~\ref{fig:6} presents the temperature
dependent shape of the octahedra in \caruoen (the slight deviations
between the high resolution and high flux results must be attributed 
to the insufficient extent in Q-space of the latter data combined with
the smaller orthorhombic splitting; nevertheless there is no doubt that these 
results show the correct tendencies). In \caruoen ~ too
we find the discontinuous change at T$_S$ : an increase in the
in-plane bond lengths and a shrinking of the Ru-O(2)-distance. But
in contrast to the stoichiometric compound the octahedron remains
slightly elongated along c till the lowest temperatures. Also the
octahedron edges parallel to the a,b-plane
present the same discontinuous changes as
S-\caruo\ but not the continuous stretching along b in the Pbca
phase.
In particular, the octahedron remains elongated along the a-axis,
i.e. perpendicular to the tilt axis. 
Due to the larger tilt this elongation may no longer over-compensate
the lattice shrinking in the a-direction, the orthorhombic lattice 
is therefore longer along b.

In \caruoen\ we also find the jump of the tilt angles at T$_S$
which remain almost constant in the Pbca phase. The transition in
\caruoen\ may be characterized as being identical to the one in
the pure compound. 
However, as there are only
minor changes of the structural parameters for \caruoen\ below the 
phase transition, one may resume that
the structure observed in the pure compound
slightly below the transition, is stable to the lowest temperatures
for the Sr-concentration x=0.1; positional and lattice parameters
of S-\caruo\ at 295\, K and \caruoen\ at 10\, K are almost
identical, see Table \ref{tab:1} and ref. \cite{7}.

\begin{figure}[!ht]
\begin{center}
\includegraphics[width=0.6\columnwidth,clip]{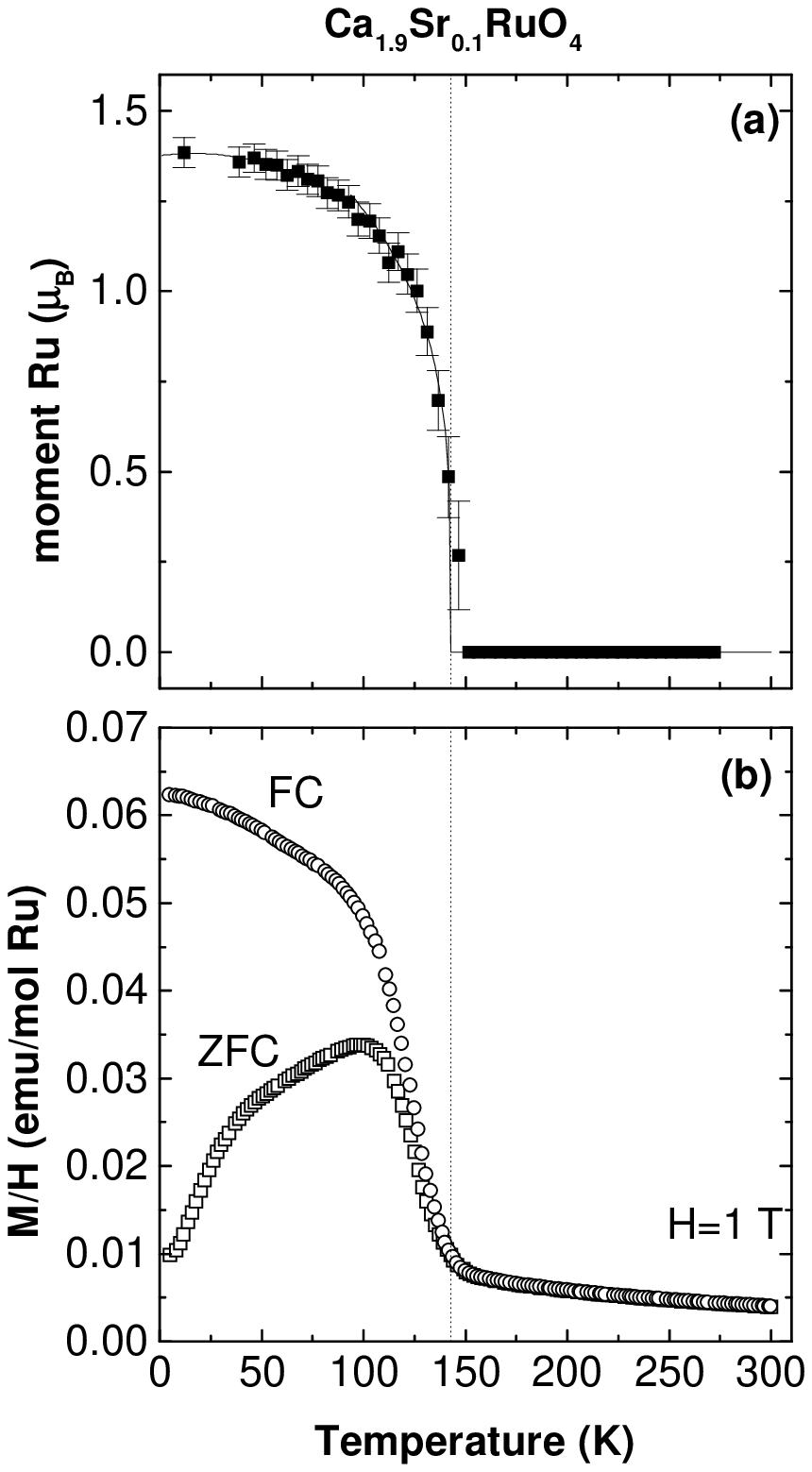}%
\caption{(a) Temperature dependence of ordered magnetic moment obtained by
refinement of the B-centered antiferromagnetic structure of
La$_2$NiO$_4$-type in \caruoen ;
(b) temperature dependence of the field-cooled and zero-field-cooled
susceptibility measured on a part of the sample.
All dependences were recorded upon heating.} \label{fig:7}
\end{center}
\end{figure}

\caruoen\ also exhibits antiferromagnetic order below T$_N$=143\,
K, see Fig.~\ref{fig:7}. There is only one magnetic transition
observed in neutron diffraction, and the magnetic arrangement is
the B-centered one of La$_2$NiO$_4$-type similar to the observation
in O-\caruo\ \cite{7}. One may note that the direction of the
ordered moment in \caruoen\ remains the b-axis, though the shape of the
octahedron is different to that of the pure compound. 
The tilt-direction seems hence to be the element determining the
spin-direction and not the stretching of the octahedron, which
underlines the importance of anti-symmetric coupling parameters.

In Fig.~\ref{fig:7} we compare the ordered magnetic moment and
the magnetic susceptibility  

That the antiferromagnetic transition manifests itself in the
susceptibility is not astonishing and arises from the
Dzyaloshinski-Moriya interaction. Weak ferromagnetism is much
stronger in the ruthenates than in the cuprates due to the
stronger spin-orbit coupling.
The low temperature susceptibility in \caruoen\ is even higher
than the values reported for La-substituted Ca$_2$RuO$_4$ \cite{cao-2000},
therefore, one may assume that magnetic order in these samples
too is mainly antiferromagnetic in nature. 
Nakatsuji et al. have observed also the
electronic  transition associated with an
upturn in the resistivity in a series of  samples of \casrruo\ 
in the Sr-concentration range up to x=0.2 \cite{nakatsuji-sces}.
These observations show that the structural transition 
triggers the magnetic order and the insulating behavior,
however the details of the coupling between structural 
magnetic and electronic transition for the highly Sr-doped 
samples need further clarification.  
The analysis of the existing data obtained 
in a cooling cycle is hampered by the fact that the down-strike
structural transition is below the intrinsic T$_N$ in the S-Pbca phase;
single crystals, however, shutter at the transition 
and therefore resistivity measurements upon heating become
difficult to interpret.

\subsection{Structure of \bcasrruo\ with x=0.2, 0.5  and x=1.0}

Upon further increase of the Sr-concentration the metal-insulator
transition is suppressed slightly above x=0.15
\cite{nakatsuji-sces}. The powder-sample of \caruoea\ studied here
does not exhibit the anomalous resistivity increase, 
nor does the one with x=0.5.

\begin{figure}[!ht]
\begin{center}
\includegraphics[width=0.9\columnwidth]{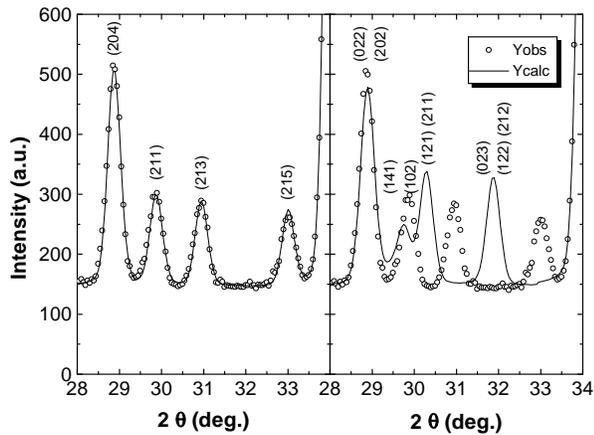}%
\caption{Comparison of the powder diffraction patterns calculated
for the rotational distortion in space group I4$_1$/acd
(c$\sim$ 24\,\AA; left) and in space group Pbca (c$\sim 12$\,
\AA; right). The calculated patterns are compared to high
resolution data observed for \caruoef .} \label{fig:8}
\end{center}
\end{figure}

The high resolution powder patterns on \caruoea\ indicate a
mixture of the  L-Pbca phase with a second phase whose space
group was identified as being I4$_1$/acd. Also the sample with
still higher Sr-content, \caruoef , was found in this symmetry. In
space group I4$_1$/acd the octahedra are rotated around the
c-axis, however, with a stacking different to that in Pbca; in
addition the tilt of the octahedra is not allowed. In I4$_1$/acd
the RuO$_2$-planes separated by $12.5\,$\AA ~ are distorted with
an opposite phase which yields a doubling of the c-parameter. This
symmetry is well known from related compounds like Sr$_2$IrO$_4$
\cite{sriro,huang} or Sr$_2$RhO$_4$ \cite{srrho}. The observation
of a phase with only a rotational distortion agrees furthermore to the
analysis of the phonon dispersion in \srruo : it was reported that only
the rotational mode is close to instability \cite{struct-inst}.
By neutron diffraction the I4$_1$/acd and Pbca phases may be
easily distinguished, since the superstructure reflections related
to the rotation occur at different l-values due to the distinct
stacking, see for example the positions of the (2 1 l)-reflections
shown in Fig.~\ref{fig:8}.

As can be deduced from the doubled c-parameter in the I4$_1$/acd
phase, a continuous transition between these phases is not
symmetry-allowed. The volume fraction of the I4$_1$/acd phase in
\caruoea\ is small and upon cooling it decreases from $21\,$\% at
room temperature to $12\,$\% at 11\, K. The insertion of the
smaller Ca enhances the internal mismatch which drives rotational
as well as any tilt transition upon cooling. Therefore one may
understand that the Pbca phase becomes more stable at low
temperature. The Pbca phase in \caruoea\ presents pronounced
structural changes upon cooling. It is found that the free tilt in
the Pbca-symmetry increases with a slight reduction of the
rotation. Also this behavior seems to be the natural consequence
of the higher Ca concentration.

A single crystal of composition x=0.2 showed
a different behavior: a single phase I4$_1$/acd structure at room
temperature followed by a tilt transition upon cooling. Since the
transition between I4$_1$/acd and Pbca is of first order, it
appears likely that the structure of samples close to the phase
boundary depends on their real structure. On the
3T.1-diffractometer the intensities of different reflections were
followed as function of temperature for the single crystal of
\caruoea. At low temperature several reflections -- for example (3
0 4) in I4$_1$/acd notation -- appear which are forbidden in the
I4$_1$/acd symmetry. The transition is continuous with only a
minor first order contribution close to the transition. The type
of the observed superstructure reflections clearly indicates that
the low temperature phase possesses a non-vanishing tilt.
However, the detailed analysis of this structure has to be performed by
neutron diffraction on a four-circle diffractometer.
Interestingly the tilt-distortion exhibits a different period 
along the c-axis, i.e. just one c-parameter 
(c$\sim$ 12\,\AA) whereas the period of the rotation is 2c. 
In general the tilt-phonon-modes show a finite
dispersion fixing the stacking sequence, due to the
displacements of the apical oxygens \cite{struct-inst}.
The interaction of the two order-parameters corresponding to different
propagation vectors, $q_1=(0.5,0.5,0)$ and $q_2=(0.5,0.5,0.5)$, 
might be an interesting problem; it certainly explains
the first order phase transition occuring near x=0.2 as a function 
of Sr-content.
The stronger tilt near x=0.2  forces the rotational distortion, 
whose stacking sequence is much less defined \cite{struct-inst}, 
into the same one-c-period.
For the following we conclude that for a Sr
concentration close to x=0.2, the structure needs a tilt distortion
either in a sub-group of I4$_1$/acd (single crystal) or in
Pbca (powder).
The tilt distortion causes an elongation of the octahedron basal
plane, as can be seen in Table \ref{tab:1} and as it discussed above.
Although the tilt may explain this elongation in a purely
structural way, the elongation should influence the
electronic band structure since the degeneration of the d$_{xz}$- and 
d$_{yz}$-bands is lifted, see below.

The high resolution patterns obtained on \caruoef\ clearly
indicate an unique I4$_1$/acd phase. There is only little temperature 
dependence in the structure of \caruoef\ as it is shown in Table \ref{tab:1}. 
In the I4$_1$/acd-structure, 
there are only three positional parameters, $z$-Ca, $z$-O(2)
and x-O(1) and only the latter changes significantly related to the
increase of the octahedron rotation angle by about $3\,\%$
between 295 and 11\, K, from 12.43(3)\grad\ to 12.78(3)\grad\ . In
the high flux patterns, no indication for long range magnetic
order has been observed yielding an upper boundary for a long
range ordered magnetic moment of 0.15$\, \mu_B$.

\begin{figure}[!ht]
\begin{center}
\includegraphics[width=0.9\columnwidth]{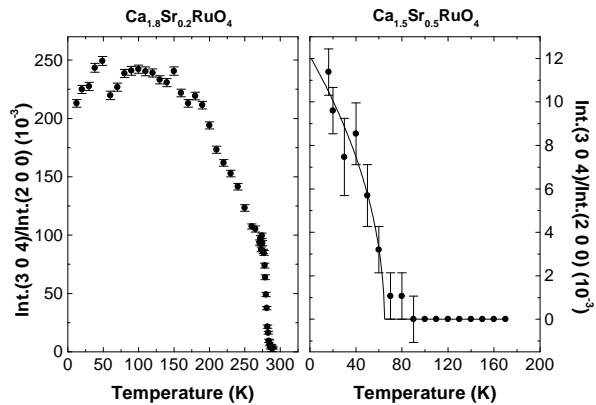}%
\caption{Temperature dependence of (3 0 4) (I$4_1$/acd notation)
superstructure reflection intensity scaled to that of the (200) fundamental reflection in
two single crystals of compositions \caruoea\ (left) and \caruoef\ (right).}
\label{fig:9}
\end{center}
\end{figure}

A small single crystal of \caruoef\ has been studied on the 3T.1
thermal neutron diffractometer as function of temperature. The
temperature dependence of the superstructure intensity shown in
Fig.~\ref{fig:9}, indicates a second order phase transition
similar to the one observed for x=0.2. The smaller Ca-content
causes a reduced transition temperature 
and a weaker tilt at low temperature,
as can be deduced from the smaller relative intensity observed in
this crystal, see Fig.~\ref{fig:9}. In the powder sample of the
same composition, x=0.5, we find only a weak anomaly in the c-axis
parameter, observed at 65\, K. The c-parameter should be quite
sensitive to an octahedron tilt, since the  projection
of the octahedra along c varies with the cosine of the tilt angle. As
may be seen in Fig.~\ref{fig:9}, the superstructure reflections
observed on the single crystal are not strong enough to be
detected by powder neutron diffraction.

The compound with x=1.0 was studied at room temperature and at 11\,
K by high resolution diffraction. We find at both temperatures a
pure I4$_1$/acd phase, however, there is evidence for disorder in
the rotation scheme. The refinement was significantly improved by
splitting the O(1)-position into two sets of positions 
(x,$|$x$|$+$\frac{1}{4}$,$\frac{1}{8}$), with x=$\pm\delta$
similar to the observation in \cite{huang} for Sr$_2$IrO$_4$. 
The average rotation
angle is temperature independent, 10.80(3)\grad\ at room temperature and
10.75(3)\grad\ at low temperature. Assuming that the square of the
rotation angle varies linearly with the concentration, one would
estimate the critical concentration for the appearance of the
pure rotation distortion near x=2.5.
From this consideration one would expect pure Sr$_2$RuO$_4$
to exhibit the same rotation in obvious contradiction to the
observation that it remains undistorted till low temperature
\cite{struc-sro}.
For a Sr concentration of x=1.5 we already do not find any long
range rotation distortion order in the powder sample.
However, there is sizeable diffuse scattering -- strong enough
to be detected by powder diffraction -- indicating that the rotation
distortion still exists on a short range scale.
The single crystal diffraction studies on the pure Sr$_2$RuO$_4$ 
\cite{struc-sro},
which are more sensitive to diffuse scattering than powder
diffraction, did not reveal any indication for a local
rotational distortion. It seems interesting to study
whether the existence of the local rotational disorder may be related
to the worse metallic properties of the Ca-doped samples.

\subsection{Phase diagram of \bcasrruo}

\begin{figure}[!ht]
\begin{center}
\includegraphics[width=0.9\columnwidth]{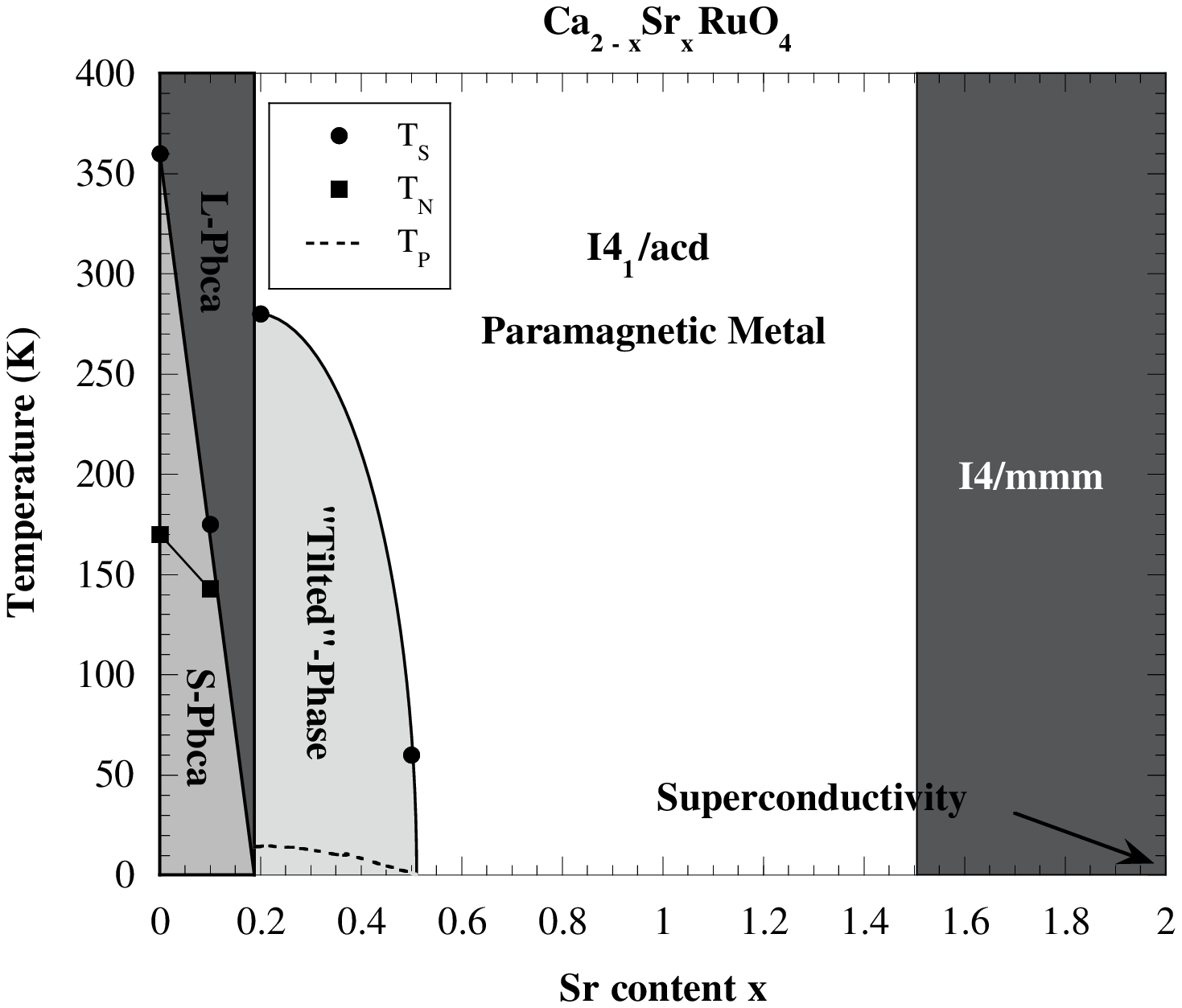}%
\vskip 0.1 cm
\includegraphics[width=0.8\columnwidth]{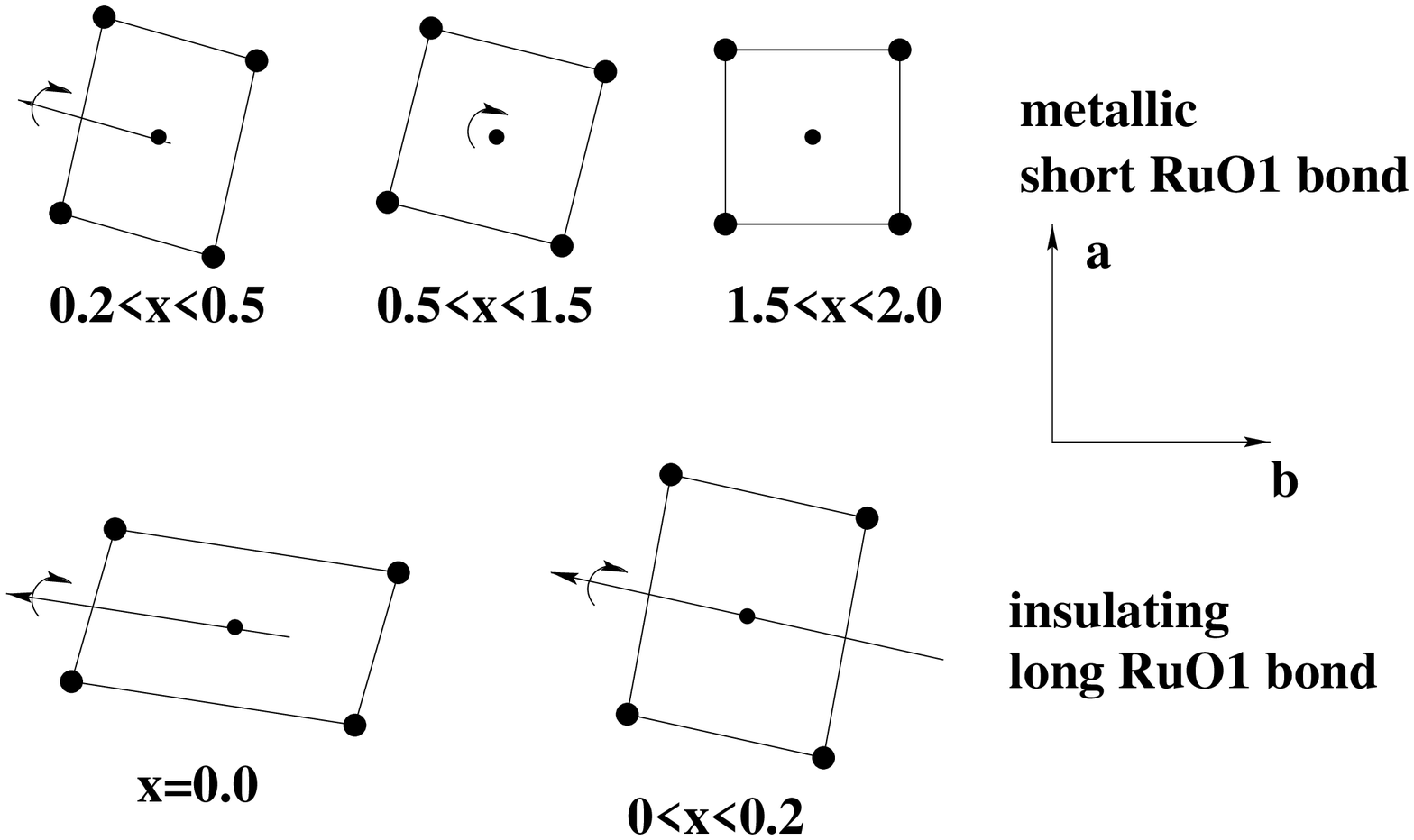}%
\end{center}
\caption{Phase diagram of \casrruo\ including the different
structural and magnetic phases and the occurrence of the maxima in
the magnetic susceptibility \cite{nakatsuji-sces}.
In the lower part, we schematically 
show the tilt and rotation distortion of the octahedra (only the basal
square consisting of the Ru (small points) and  the O(1) (larger points)
is drawn) together with the elongation of the basal planes.
} \label{fig:10}
\end{figure}

\begin{figure}[!ht]
\begin{center}
\includegraphics[width=0.6\columnwidth]{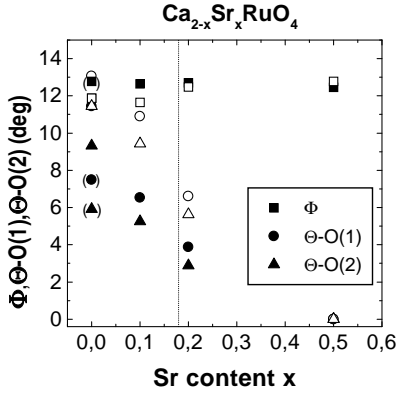}%
\caption{The Sr concentration dependence of the tilt $\Theta\text{-}O(1)$ and $\Theta\text{-}O(2)$ and rotation angle $\Phi$ in \casrruo. The filled symbols denote the results obtained at T=300\,K (those with brackets are obtained at T=400\,K) and the open symbols those at T=10\,K, the 
dashed line indicates the critical concentration below
which one observes the insulating S-Pbca phase at low temperature.} 
\label{fig:12}
\end{center}
\end{figure}

According to the studies presented above the structural and
magnetic phase diagram of \casrruo\ is rather complicated, a
schematic picture is given in Fig.~\ref{fig:10}, which also
presents the results of the magnetic studies by S. Nakatsuji et
al. \cite{nakatsuji-sces}. 
In addition the tilt and rotation angles together with
the deformation of the octahedron basal plane are depicted.
At low temperature and for decreasing
Sr-concentration, one passes from the undistorted K$_2$NiF$_4$
structure in pure \srruo, space group I4/mmm, to a simple rotation
I4$_1$/acd in agreement to the low lying rotation mode in the pure
compound \cite{struct-inst}. An estimated boundary
is included in the diagram in Fig.~\ref{fig:10}, though the
transition is found to exhibit order-disorder character.
For a Sr-concentration of x=1.5 only diffuse scattering
representative of a short range rotational distortion is
present. 
For x=1.0 the rotational angle already amounts to 10.8\grad\ ;
the rapid suppression of the structural distortion in Sr-rich samples
appears to be extraordinary it might hide some further effect.
For much smaller Sr-content,
near x=0.5, a combination of rotation and small tilt is found.
This is realized
either in a subgroup of I4$_1$/acd or in Pbca. Further
decrease of the Sr-content leads finally to the combination of the
rotation and the large tilt in the S-Pbca phase.
The Sr-dependence of the tilt and rotation angles is resumed in
Fig.~\ref{fig:12}.
Most interestingly all these
structural transitions are closely coupled to the physical
properties.

The purely rotational distortion should be related to the c-axis
resistivity since it modifies the overlap of the O-orbitals in
c-direction. This rotation phase becomes unstable against
the tilt for Sr concentrations lower than 0.5, since in the single
crystal with x=0.5 only a minor distortion has been observed,
which remained undetectable in the powder sample. For the
Sr-concentration of x=0.2 we already find tilt angles of about 7\grad\ 
at low temperature by powder neutron diffraction. Near
x=0.5 there is hence the quantum critical point of the continuous tilt
transition which coincides with a maximum in the low temperature
magnetic susceptibility. For x=0.5, Nakatsuji et al. report a low
temperature magnetic susceptibility about 100 times larger than
that of pure \srruo\ \cite{nakatsuji-sces}. This suggests 
that the low-lying tilt modes are strongly coupled
to the magnetism. 
This interpretation is further supported by the
fact that in all magnetically ordered structures, x=0.0, x=0.1 and 
in O-Ca$_2$RuO$_4$, the spin-direction
is parallel to the tilt axis in spite  of a different octahedron shape
as it is schematically drawn in figure 10.
Further decrease of the Sr-content 
below x=0.5 stabilizes the tilt
and causes a maximum in the temperature dependence of the
susceptibility at T=T$_P$, indicated in Fig.~\ref{fig:10}
\cite{nakatsuji-sces}. T$_P$, however, does not coincide with the
structural transition from I4$_1$/acd to the tilted phase  
but is much lower. 
We speculate that the susceptibility maximum 
arises from an increase of antiferromagnetic fluctuations
induced by the tilt.

There is another anomalous feature in the 
temperature dependent susceptibility of samples with
0.2$<$x$<$0.5 : Nakatsuji et al. find a strong anisotropy
between the a and b-directions of the orthorhombic lattice 
\cite{nakatsuji-sces}.
The details of the tilt and rotation distorted structure  
observed for single crystals in this Sr-range
have still to be clarified, but at least for x=0.2
the crystal structure is known, see Table \ref{tab:1}. 
The complex structure combining the one-c tilt with
the two-c rotation should be identical as far as a
single layer is concerned.
In relation to the magnetic order in the insulating compounds
it appears again most likely that the tilt axis, which is  parallel b, 
is the cause of the huge anisotropy.
However, there might be an additional  more complex effect related to the 
band structure, which is described for Sr$_2$RuO$_4$ for example
in reference \cite{mazin}.
Three bands are crossing the Fermi-level a quasi-two-dimensional
band related to the d$_{xy}$-orbitals and quasi-one-dimensional
bands related to d$_{xz}$- and d$_{yz}$-orbitals.
The elongation of the octahedron basal plane
will lift the degeneracy between d$_{xz}$ and d$_{yz}$-orbitals
(in order to demonstrate this,
one may choose an orbital set rotated around the
c-axis by 45\grad) and might via this mechanism
enhance the anisotropy.

One might also argue that the splitting of the Ru-O(1)-distances 
which is allowed in L-Pbca is related to a Jahn-Teller
effect \cite{nakatsuji-mag}. However, there is no clear experimental evidence for 
such a splitting, fits of similar agreement  
may be obtained when constraining the
bonds to equal length (R$_{wp}$ increases in all
cases by less than 0.02\%). 
Furthermore, these bonds are not pointing along 
the direction of the observed anisotropy but along the diagonals of the
orthorhombic lattice, and, finally, short and long bonds would be alternating
in the sense that a Ru-O-Ru path always consists of a short and a long bond. 
Hence, even though it is possible that Ru-O(1) distance splitting 
might be related  to orbital ordering, 
it could not explain the observed magnetic anisotropy.

For Sr concentrations lower than 0.2 we find the Pbca symmetry
and the first order phase transition leading to the insulating S-Pbca phase.
The metal-insulator transition is hence observed as a function of
concentration as well as a function of temperature.

\begin{figure}[!ht]
\begin{center}
\includegraphics[width=0.6\columnwidth]{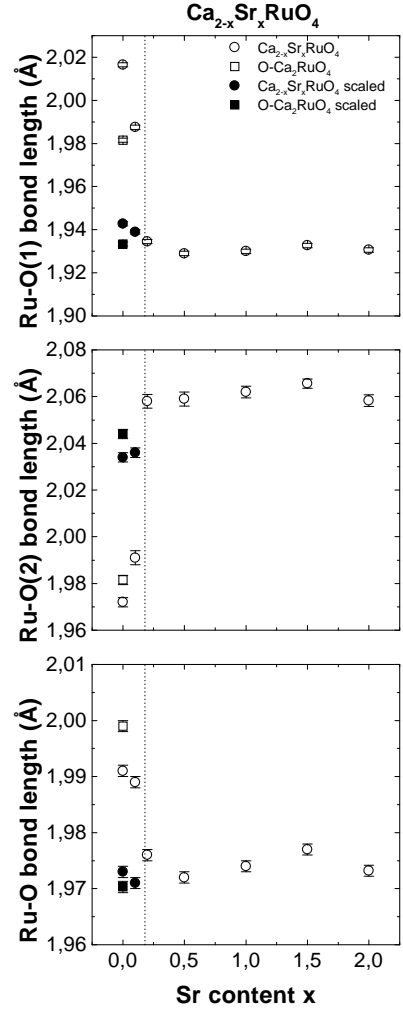}%
\caption{Composition dependence of the Ru-O(1)-, Ru-O(2)- and 
averaged Ru-O-bond distances in
\casrruo ~ at 10\, K open symbol.
Filled symbols were obtained by extrapolating the distances
obtained in the high temperature metallic phase to 10\ K using 
the thermal expansion of Sr$_2$RuO$_4$; the dashed line designates
the critical concentration for the metal insulator transition.} 
\label{fig:11}
\end{center}
\end{figure}

In the Sr range 0.2$\leq$x$\leq$2 all samples are found to be
metallic at low temperature and there is little variation of the
octahedron shape and in particular of the in-plane Ru-O bond
lengths when compared to that observed in \srruo\
\cite{struc-sro,chmaissem}. The concentration dependent octahedron
shape is shown in Fig.~\ref{fig:11}. Metallic 214-ruthenates
appear to be identical at least concerning their Ru-O-bond distances.
The minor deviation near x=0.2 may be explained by the relaxed
constraint in the tilted structure. For smaller Sr-content the
metal insulator transition occurs, but one may still compare to
the Ru-O(1)-bonds in the metallic high temperature phase. For this
purpose we have scaled the high resolution results in the metallic
phase of the three samples presenting the metal insulator
transition, stoichiometric \caruo, O-\caruo\ and \caruoen, by
the thermal expansion of \srruo\ \cite{chmaissem,comment}. Also
these values are comparable to pure \srruo, which demonstrates the
equivalence of the metal insulator transition as function of
temperature and as function of concentration.

Concerning the metal insulator transition it appears necessary to
separate two effects. In all samples exhibiting an insulating
low temperature phase there is an increase of the in-plane
bond lengths and a reduction of the Ru-O(2)-distance
accompanied with the increase of the tilt angle.
These effects have to account for the non-metallic behavior.
Within a Mott-scenario \cite{mott} one may phenomenologically
explain the effect, since both tilting and increase of the
in-plane distances should strongly reduce the band-width. In
particular the band corresponding to the $d_{xy}$-orbital should
become more localized and lower in energy. Assuming that there is
a sizeable Hubbard $U$ in these ruthenates, the $U/W$-ratio might then
pass above one explaining the nonmetallic behavior.

In the stoichiometric compound the transition is not
complete immediately below T$_S$; instead the octahedron becomes
flattened mainly due to the elongation of the basal plane along
the b-axis which is the direction of the spins in the
antiferromagnetic ordered structure. We do not think that this
rather peculiar behavior can be explained by simple structural
arguments. It appears likely that the spin-orbit coupling in the
non-metallic phase causes the low temperature structural
anomalies. The reason why similar effects do not occur in
O-\caruo\ and in \caruoen\ might be found in their remaining
itinerant character. Spin-orbit coupling also forces the spin
direction parallel to the tilt-axis, which will strongly reduce
the weak ferromagnetism along the c-direction. 
Due to the flattening the d$_{xy}$-orbital
should be well separated in energy and be filled, see also the discussion 
in \cite{nakatsuji-mag}. 
The remaining two electrons occupy the d$_{xz}$ and d$_{yz}$-orbitals 
whose degeneration is lifted by the elongation of the octahedron basal 
plane along b by spin-orbit coupling or by
a Jahn-Teller effect \cite{7}.

\section{Conclusion}

The phase diagram of \casrruo\ shows a variety of different
structural magnetic and electronic phases. The distinction between
metallic and insulating compounds appears to arise from an
enhanced in-plane Ru-O bond distance and a larger tilt. 
This behavior
suggests the interpretation as a Mott-transition
related to the lowering in energy of the d$_{xy}$-band.
Samples with strongly flattened octahedra are driven non-metallic
by the structural transition, whereas in a sample with a reduced
flattening the metal-insulator transition is close to 
the magnetic ordering. The local character of the moments
in the insulator induces an additional structural distortion via
spin-orbit coupling.

In the part  of the phase diagram where samples
stay metallic to low temperature,
first a rotation and second a tilt distortion develops
upon increase of the Ca-content.
This should be considered as the purely structural
consequence of the smaller ionic radius of the Ca.
The observation of the maxima in the temperature dependence of the magnetic
susceptibility only for samples presenting an octahedron tilt,
and the fact that the maximum low temperature susceptibility is
found near the composition where the tilt distortion vanishes,
clearly indicate strong magneto-elastic coupling.

The tilt seems to play a key role in the magneto-elastic coupling 
since in all magnetically ordered structures the spin direction 
is parallel to the tilt even though the shape of the octahedron
is rather distinct. Furthermore, in the intermediate metallic
region, 0.2$<$x$<$0.5, which exhibits a tilt distortion,
a strong anisotropy is found for the 
susceptibility parallel and perpendicular to the tilt axis.
Due to purely structural constraints the tilt leads to an
elongation of the octahedron basal plane, which may lift
the degeneracy of the d$_{xz}$ and d$_{yz}$-orbitals.

The magnetic and electronic degrees of freedom in the
214-ruthenates are hence closely coupled to each other and also to
the lattice. Whether one of these couplings -- and which one -- is
responsible for the occurrence of superconductivity in \srruo,
however, remains an open question.

\begin{acknowledgments}
We wish to acknowledge D.I. Khomskii, I.I. Mazin and P. Pfeuty
for stimulating discussions.
\end{acknowledgments}



\begin{thebibliography}{}
\bibitem{comment}The scaling by the thermal expansion of pure
\srruo\ is not fully justified, since the temperature dependent
tilt and rotation should alter the thermal expansion. This and the
insertion of additional oxygen should explain the minor
deviations.
\end{thebibliography}
\end{document}